\documentstyle[aps,twocolumn,epsfig]{revtex}

\def\Esf{E_{\rm sf}}
\def\Htile{{\cal H}_{\rm tile}}

\begin{document}

\title{Total-energy-based structure prediction for d(AlNiCo)}

\author{
C.L. Henley$^1$, 
M. Mihalkovi\v{c}$^{2,3}$, 
and M. Widom$^4$}
\address{
$^1$ Department of Physics, Cornell University, Ithaca NY  14853 \\
$^2$ Institute fur Physik, Technische Universit\"{a}t Chemnitz, 
D-09107 Germany \\
$^3$ Institute of Physics, Slovak Academy of Sciences, Bratislava, Slovakia \\ 
$^4$ Department of Physics, Carnegie-Mellon University, Pittsburgh PA  15213 \\
}
\date{\today}

\maketitle

\begin{abstract}
One may predict a quasicrystal structure starting
from electrons and quantum mechanics, as approximated by 
interatomic pair potentials calibrated with ab-initio total-energy 
calculations, combined with 
the experimentally known composition and lattice constants.  
Here we report our progress 
on the ``basic Ni'' decagonal phase $d$(Al$_{70}$Ni$_{21}$Co$_{9}$.
Atomic configurations are represented  as decorations of 
(possibly) random tilings.  
Our method was Monte Carlo simulation using both lattice-gas hops 
by atoms and tile-flip rearrangements, eventually followed by molecular 
dynamics and relaxation of the atom positions. 
Initially allowing the greatest 
freedom of atom positions, we observed nearly deterministic structural 
rules and enforced these as constraints involving larger tiles; 
this procedure was repeated at the next level of modeling. 
In crude and preliminary form, the
effective Hamiltonian for tile-tile interactions is known, which is needed 
for further simulations to infer the long-range order. 
Our atomic arrangements in 
the 20~\AA~ decagonal cluster are compared with three structure
models based on recent experiments.~

\end{abstract}

\section {Introduction} 

This paper reports on early results of a program to determine
the structure of a quasicrystal -- specifically $d$(AlNiCo) --
strictly from energetics, without use of diffraction data.
Almost all of the work was performed by M.M. and M.W., with their
coauthors,  and is reported in Ref.~\onlinecite{alnico}, 
but the present account is
tilted towards the particular interests of C.L.H. 

We have focused on the thermodynamically stable and highly ordered 
``Basic Nickel'' subphase of 
decagonal $d$(AlNiCo)~\cite{basicNi,PhaseDiagram}, 
of composition approximately 
Al$_{0.70}$Ni$_{0.21}$Co$_{0.09}$.
Our ultimate objective is to {\it predict} the structure on the basis
of total energy; though the gross aspects of the structure are clear from 
diffraction, there are many ambiguous or controversial details, 
and there is no understanding for the quasicrystal's special
stability and simplicity at this composition. 
In particular, we wish to identify the sites of the
transition metal (``TM'') components Ni and Co,  
which are indistinguishible to (ordinary) X-rays 
or electrons~\cite{takakura,cervellino,yan,abe,FN-NiCo}.

In a sense, the most difficult problem of crystal chemistry is
to predict which structure a given composition will adopt, even 
if one has an exact and tractable Hamiltonian for the total energy
(e.g. by pair potentials) -- for
this necessarily involves a comparison with an {\it infinity}
of possible structures.  The only irreproachable method is
a mathematical proof, which is feasible in rare cases with 
short-range interactions, e.g. close-packing of hard spheres, 
or the two-dimensional ``binary-tiling'' quasicrystal 
(a toy model)~\cite{Binary}. 
A naive, brute-force
approach would be to cool (in simulation) from the melt, and see
which structures emerge; but this is prone to fail for a complicated
material, since the accessible times are so short that it will
get stuck in a glassy disordered configuration.  
Even when a quasicrystal emerges, as with 
Dzugutov's toy potentials~\cite{dzugutov}, 
the system may have found a merely metastable ordered
state for kinetic reasons: indeed, the stable phase in this case
turned out to be a simple bcc packing~\cite{rothDzu}.

\section {Decoration models}

We represent the quasicrystal structure as a decoration of
{\it disjoint} tiles. 
There are several reasons for doing so, rather than as a irrational
cut through a five-dimensional hypercrystal.  Based in real space, this
representation is easier to visualize and somewhat more tractable, 
technically, than the hypercrystal approach. For example, shifts of the
atoms from ideal (tiling-vertex) sites are parametrized by just a finite set 
of real numbers, like the coordinates in a crystal unit cell. 
A decoration description -- particularly in the decagonal case -- naturally
lends itself to a hierarchy of supertilings, which are presumably
involved in the modulations which distinguish some of the
subphases in the $d$(AlNiCo) phase diagram~\cite{PhaseDiagram}.
Furthermore, most hyperspace structures which have been defined using
discrete acceptance domains, e.g. Ref.~\onlinecite{takakura}, 
can be translated into tiling-decoration language. 

Decoration models are uniquely suited to a correct modeling of
chemical disorder (including vacancies) in a quasicrystal. 
Of course, a hyperspace fit to Bragg data can incorporate  mixed 
occupation of a site type, but it is highly implausible (from the
viewpoint of the structural energy) that
the occupations of neighboring sites vary {\it independently} of
each other.  
If the occupations are strongly correlated, then the mean structure
may differ significantly from any particular real structure;
this matters e.g. because the electronic structure is
quite sensitive to the TM-TM 
contacts in $d$(AlNiCo)~\cite{krajci}.
The tiling decoration is a very convenient way to
account for such occupation correlations, by ascribing the randomness
to the tiles
and not the decoration. 
The ultimate criterion of a decorated model -- fixing e.g.
the appropriate size of tile -- is that the 
ensemble of allowed tilings corresponds one-to-one with the 
ensemble of low-energy arrangements in an atomistic model. 

A decoration description replaces the independent real coordinates 
of many atoms by tiling degrees of freedom, which are discrete and
many times fewer. When the decoration is deterministic and disjoint
(each atom bound to a particular tile or tile-related geometric 
object), then the atoms' interaction energies can be expressed as
a function solely of the tile configuration, 
called the ``tile Hamiltonian'' $\Htile$.
This facilitates faster Monte Carlo simulations, 
on the comparatively few tile degrees of freedom, 
in order (i) to reveal supertilings in the process of discovering
the atomic structure, as in this work; 
(ii) to predict diffuse scattering, or (iii) eventually to measure the 
phason elastic constants (which govern the diffuse wings surrounding
Bragg peaks.)  
Armed with a reliable $\Htile$, one can also simulate three-dimensional
samples to obtain quantitative estimates of the perfection of the 
long-range order and the strength of diffuse diffraction. 

This approach may also be the easiest way to resolve, for a particular
stable quasicrystal, whether or not it is entropically stabilized. 
It is stabilized by energy favoring a quasiperiodic state
if $\Htile$ happens to implement the Penrose `matching rules'.
That {\it almost} occurs in $d$(AlCuCo), as modeled
with potentials like those used in the present work~\cite{cockayne}. 

Our challenge, then, is to find the decoration rule and the 
tile-packing constraints without any bias, apart 
from the above-mentioned assumptions about the lattice constants. 
A caveat to keep in mind is that the results will
probably be quite sensitive to the exact composition and 
number density.

\section {Inputs: potentials and constraints}
\label{sec-inputs}

The main input to our calculations is a set of atom-atom pair potentials
$V_{ij}(r)$ for species $i$ and $j$, six distinct
functions in the case of a ternary alloy. 
They are given by Moriarty's 
``generalized pseudopotential theory'' (GPT)~\cite{GPT-I,GPT-II,GPT-IV}, 
a systematic expansion of the total energy
as a sum $E_0+E_1+E_2+\ldots$, where the $E_n$ term depends on
$n$ atoms,  but we use only the $n=2$ terms. 
The first-neighbor well of the raw $V_{\rm TMTM}$ is unphysically
deep (which would be canceled by $E_3$ in the systematic theory).
Therefore our $V_{\rm TMTM}$ potentials were empirically modified, 
by fitting a short-range repulsive correction so as to match 
all the forces in a full density-functional calculation 
on a small (50 atom) approximant of $d$(AlNiCo)~\cite{GPT-IV}. 
There is strong support that our $V_{ij}(r)$ have
the correct $r$ dependence and relative strength, 
since they predict the Al-Co and Al-Ni 
phase diagrams pretty well~\cite{GPT-II}
(as a function of concentration at $T=0$);
however the ternary Al-Ni-Co phase diagram was not attempted. 
On the basis of the simulated melting temperature and
phonon spectra,  we do suspect that our potentials
are $\sim$30\% stronger than the reality.

Like the similar Al-TM potentials of Phillips, Zou, and Carlson, used
earlier in Refs.~\onlinecite{cockayne} and \onlinecite{decoII}, 
the GPT potentials depend implicitly on the
net valence electron density, and
exhibit strong ``Friedel oscillations'' as a function of distance
(tails decaying as $\cos(2 k_F r + \delta)/r^3$). 
For example, $V_{\rm AlTM}(r)$ and $V_{\rm TMTM}(r)$
show four prominent minima in the range $r \leq 8.5$~\AA;
such minima are quite important in
deciding the structure~\cite{decoII,FN-HumeR}.

Notice that V$_{\rm AlAl}(r)$ is practically zero
after its hard core radius $\sim 2.9$~\AA~:
there is not even a nearest-neighbor well, 
just a shoulder on the tail of the hard-core. 
The V$_{\rm AlTM}$  potentials have very deep nearest-neighbor wells:
this creates the illusion of a TM-TM repulsion, but really
TM-TM neighbors are avoided mainly to 
make more room for Al in the TM coordination shells. 
However, at roughly 30\% TM content, some TM-TM contacts
are unavoidable; they are all Ni-Ni, since 
$V_{\rm AlCo}$ has a deeper well than V$_{\rm AlNi}$.
Finally, the $V_{\rm TMTM}$ potentials have their deepest
well at second-neighbor ($\sim 4.2$\AA) distances. 
The consequence of all this is that the TM atoms form a 
rather rigid and somewhat uniformly spaced network~\cite{FN-whystable},
while Al atoms move rather freely to follow the potential wells 
or troughs created by the TM arrangement. (See Sec.~\ref{sec-cluster}
for more discussion).

Our use of these potentials imposes some limitations. 
We omitted the $n=0$ term, which depends only on the valence
electron density, and is the largest contribution to a metal's total energy.
Hence we can make valid comparisons only between structures with 
(practically) the same valence electron density. 
We cannot meaningfully predict the lattice constants.
Nor can we even expect the quasicrystal phase to be
globally stable according to our potentials, since they might
spuriously favor phase-separation.

In view of the above-mentioned pitfalls for a brute-force approach, 
we adopted as a {\it second} input
the experimental {\it lattice constants}~\cite{basicNi}:
the stacking period is $c=4.08$~\AA~ 
and the quasilattice constant in the decagonal plane is
$a_0=2.45$~\AA.
We only seek the lowest energy among arrangements on this framework. 
This highly constraining assumption
still permits a vast ensemble of possible structures. 

\section {Methods and results}

Our methods are a mix of tile-flip Monte Carlo, atom-hopping 
Monte Carlo, relaxation of atomic positions, and molecular dynamics. 
Our procedure is first to discover the favorable
low-energy motifs through Monte Carlo annealing, then to
remove unnecessary degrees of freedom, and repeat, 
producing successively more constrained models.
Having fewer degrees of freedom the latter
are much faster to simulate at low temperatures.
At the end, we can investigate the effects of letting
atoms depart from ideal sites.

The initial stage of our exploration stacks two
independent small (edge $a_0$)
Penrose rhombic tilings, in a vertical space of one
lattice constant $c=4.08$~\AA~ and
the only allowed atom sites are on vertices. 
A manageable size was 50 atoms on 72 candidate sites, initially
on a good Penrose-tiling approximant; periodic boundary conditions are 
used in all directions (and at all stages of exploration). 
The atoms -- initially
chosen to approximate the experimental density and composition of
the Ni-rich ``Basic Ni'' phase -- hop as a lattice gas on these sites.
This allows sufficient freedom for the atoms, if they ``want'', 
to adopt any of the decagonal structure models -- with stacking
period $c=4$~\AA~ -- that were hypothesized
at one time or another~\cite{hen-decag,FN-cockayne}. 
The Monte Carlo moves permit swaps of the species between two nearby sites 
(``vacant'' is a special case of species!), as well as 
``tile flips'' which reshuffle the three rhombi in a fat or 
thin hexagon in the same layer.

\begin{figure}[!ht]
\centerline{
\epsfig{file=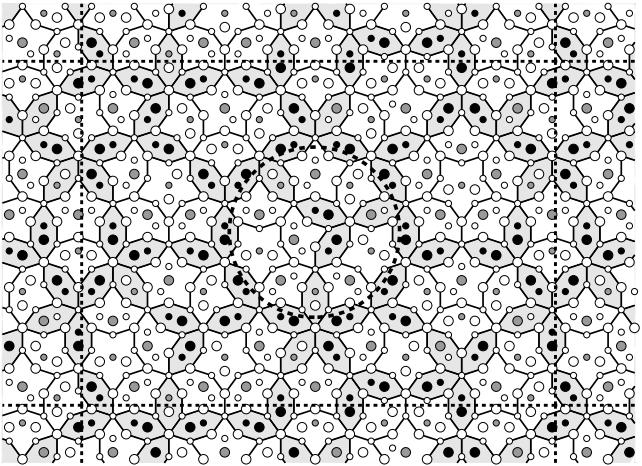,width=2.7in,angle=-90}}
\caption{Lowest-energy configuration obtained from a simulation
with variable occupancy interiors.  
Circles are colored with the
following convention: empty=Al, gray=Co, black=Ni;
the size of the symbol represents the $z$ coordinate.
Shaded small hexagon tiles, occupied by either NiNi pairs or AlCo pairs, 
form the edges of an HBS supertiling.
Dashed circle indicates an 20~\AA~ decagon cluster.
This is the same configuration as Fig.~4(b) of 
Ref.~\protect\onlinecite{alnico}.}
\label{fig-large}
\end{figure}

After this model is slowly cooled to zero temperature, 
if the initial composition was rightly chosen, 
one obtains a {\it one}-layer Hexagon-Boat-Star (HBS) 
tiling of edge length $a_0$, 
with a {\it two}-layer decoration in which 
the allowed sites lie over vertices of
the Penrose rhombi (into which HBS tiles may be decomposed) or an additional
site in each Fat rhombus. 
This decoration places Al atoms over the HBS
vertices (even and odd vertices alternate
between even and odd layers). Isolated Co atoms 
sit over each Boat and Star tile center, 
ringed by possible Al sites;
in the Star at most two Al can 
be present, out of five ideal sites in this ring. 
Particularly characteristic are the vertical zigzag chains,
with Ni in each layer (appearing as NiNi pairs in Fig.~\ref{fig-large}), 
over the interior of every Hexagon tile. 
However, NiNi may be replaced by CoAl in some places. 
Decorating a tiling with the same H/B/S ratios as the quasiperiodic
Penrose tiling yields an
ideal composition Al$_{0.700}$Ni$_{0.207}$Co$_{0.093}$ and atomic
volume 14.16~\AA$^3$. This composition coincides with the
experimental Basic Ni phase, but the number density of
atoms is at least 5\% greater than in experiment.

For the next stage of modeling, to discover larger-scale regularities, 
the small HBS tiles which emerged from the initial stage are
elevated to fundamental objects, either with a rigid decoration 
or with some atoms fixed and others forming a lattice gas. 
The allowed tile flips are reshufflings of fat hexagons 
of the underlying rhombus tiling, provided the result is
a valid HBS tiling and conserves the atom content.
Additionally the Al pair inside the Star tile can rotate among five
allowed orientations.  
From this simulation, it emerges that the
Hexagon tiles containing Ni chains only touch tip-to-tip, 
so that the angles relating them are multiples of $72^\circ$.  In fact, 
(see Fig.~\ref{fig-large}) the long axes of the Hexagons
form edges of an HBS {\it supertiling} with an 
inflated edge length $\tau^2 a_0$, where $\tau \equiv (1+\sqrt5)/2$.

\section{The 20 \AA ~ decagon cluster}
\label{sec-cluster}

We now compare our results with  well-known structure models,
organizing the discussion 
around the famous 20~\AA~ diameter 
decagon cluster, which is prominent in electron-microscope 
images and in most diffraction refinements as well. 
(The ideal decagon diameter is actually $2 \tau^3 a_0$.) 
Such decagons indeed appear in our structures~\cite{FN-fakering};
Fig.~\ref{fig-large} is centered on one of them. 
Z-contrast images, in which the intensity is a direct projection
of atoms weighted by squared atomic number, 
reveal the TM positions~\cite{yan,abe}; 
the TM (and most Al) positions proposed by
Ref.~\cite{abe} are practically the same 
as in our version of the cluster.

For a more detailed comparison, we focus on three recent structural 
studies~\cite{takakura,cervellino,yan}.
All of these, {\it and our simulation}, agree on the following details
of the projection: (1) the outer decagon (edge length $\tau^2 a_0$)
has Al on each vertex and a pair
of TM atoms on each edge,  which we identify as Ni. (2) A middle decagon
(edge $\tau a_0$) has TM atoms on each vertex, which we identify as Co. 
(3) An inner decagon (edge $a_0$) has Al on each vertex.
(4) In the center, the 10-fold symmetry is broken and a sort of
isosceles triangle is observed, with one TM (we say Co) at the unique
corner and pairs of TM (we say NiNi) at the base corners. 
The cluster is evident in Ref.~\onlinecite{takakura} on the
right side of their Fig.~7, as a combination of a Boat tile + 2 Hexagon
tiles, just as in our Fig.~\ref{fig-large}. 

\begin{figure}[!ht]
\centerline{
\epsfig{file=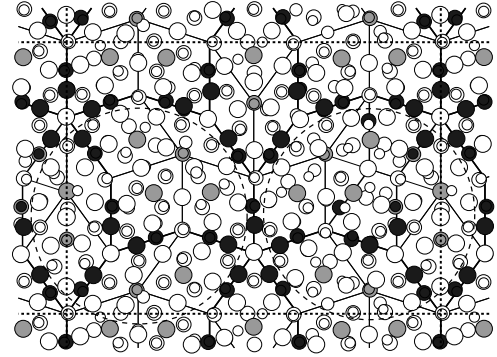,width=2.7in}}
\caption{Two bilayers (thickness $2c$), with periodic boundary conditions 
in all directions, after a tile flip in one bilayer, 
followed by molecular dynamics and relaxation to an energy minimum. 
Black lines are HBS supertile edges (length $\tau^2 a_0$).
The dashed circles are imperfect 20~\AA~ decagons --
this cell is too small to contain complete ones.}
\label{fig-relaxed}
\end{figure}

Detail (4) is somewhat controversial, since some decagon
images have non-triangular centers. Indeed, the density maps
from the refinement of 
Ref.~\onlinecite{cervellino} show {\it six} strongly TM sites at
the center of many (but not all) of the 20~\AA~ decagons.
We ascribe this to {\it stacking flips} between one layer and the 
next layer (see Sec.~\ref{sec-tileH}), 
seen in projection, as 
in the right-side decagon of Fig.~\ref{fig-relaxed}. 

The simulations we described 
up till now implicitly assumed a strict $c=4.08$~\AA~ periodicity in
the stacking direction, 
neglecting (like so much other modeling) the fact that 
decagonal quasicrystals are three-dimensional.
To obtain the configuration in Fig.~\ref{fig-relaxed},
we increased the periodicity
to $2c$, i.e. two HBS layers, which initially were the same tiling.
First a tile-flip was made in one of the two HBS layers,
then the tilings were decorated, annealed by MD, and relaxed to
an energy minimum. 

In a fifth, controversial detail of the decagons, our model 
initially disagreed with experiments in which 
every edge of the middle decagon shows a {\it pair} of Al atoms. 
These sites form vertical zigzag chains, but they cannot all be occupied since
they are separated by essentially the interlayer spacing $c/2=2.08$~\AA.
In our fixed-decoration model, then, these sites are occupied in only one
atom layer, with separation $c$,  so only one atom is visible on each edge. 
We found that, after molecular dynamics (MD) at 1000K followed by 
relaxation, some Al neighbors of the Co atoms
shifted to the middle-decagon edges, which now present Al doublets 
in projection, in agreement with experiment.  

\begin{figure}
\centerline{\epsfig{file=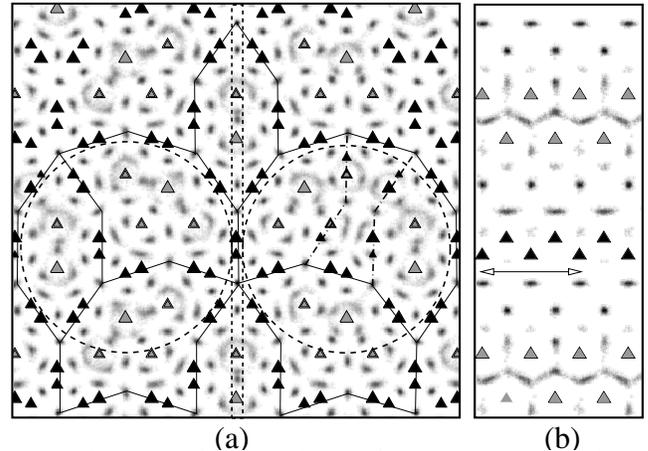,width=3.4in}}
\caption{Probability density of Al atom centers during a 
molecular dynamics simulation of the same configuration as 
in Fig.~\ref{fig-relaxed}.
TM locations are indicated by triangles, with the same coloring and 
size conventions as the previous figures.
Edges of the large HBS tiles are marked by lines.
(a) Projection onto xy plane. 
(b) Projection onto the yz plane of the narrow slab outlined
by a dashed rectangle in (a). The pattern repeats with the
z period 8~\AA.
(Note this slice actually cuts through an H tile but
not a 20 ~\AA~ cluster.)}
\label{fig-MDfuzz}
\end{figure}

To further understand what happens with these atoms, 
the time-average of the Al positions during the MD simulation
is shown in Fig.~\ref{fig-MDfuzz}. 
Some 40\% of Al atoms are rather delocalized, and would need
a highly anisotropic Debye-Waller factor in a 
crystallographic fit.~\cite{abe-phason}
The vertical projection (Fig.~\ref{fig-MDfuzz}(a)
shows, consistent with our remarks on the potentials
(Sec.~\ref{sec-inputs}), 
that each Co atom in the middle decagon is surrounded by a potential trough
in which Al atoms appear almost free to roll 
like ball-bearings~\cite{gaehler-MD}.

A slice along a vertical plane further clarifies the Al behavior:
a second type of trough extends vertically, with a zigzag shape, 
and in fact connects with the circular troughs.  In our simulations
with a cell $2c$ 
in the vertical direction, we found {\it three}
Al atoms appearing in each zigzag trough (per $2c$,  i.e. per 4 atom layers). 
Notice that the $z$ displacement of two of these atoms makes the
layer puckered, as is already known from diffraction~\cite{cervellino}, 
and implies a local cell-doubling along the $c$ axis, similar
to the $\rm Al_{13}Fe_4$ or $\rm Al_{13}Co_4$ decagonal approximants. 

The symmetry-breaking of the cluster interior
was recently predicted from energies 
by a full {\it ab-initio} calculation~\cite{yan-calc}.  
They find, from a quite different starting point than ours,
the identical arrangement of the 5 Al + 5 TM atoms (per atomic
bilayer) found at the cluster's center.
However, they rejected
the possibility of Al doublets on the middle decagon edges without
trying $z$ relaxations, nor did they address Co/Ni ordering. 
We find it striking that our mere pair potentials sufficed to
obtain the experimentally indicated form of symmetry breaking 
in the cluster center,
while using {\it no experimental input whatsoever} regarding the positioning
of Al relative to TM atoms.

\section {Tile Hamiltonian and three-dimensional stacking}
\label{sec-tileH}

To model a large, three-dimensional sample using
a tile Hamiltonian, we cannot demand perfect periodicity
in the $z$ direction. 
(One reason is that the ``entropic stabilization'' 
explanation of quasicrystal order requires an extensive entropy.)
So, we represent each bilayer of atoms by a distinct HBS tiling, 
and constrain adjacent tilings in the stack to differ 
by ``stacking flips'', 
which are exactly the same reshufflings 
that constitute our Monte Carlo tile-flips.

We also found (from simulations) that 
-- among (small) HBS tilings in which H tiles form a network
touching tip-to-tip -- the in-plane tile 
Hamiltonian is dominated by interactions between 
``H(NiNi)'' tiles (Hexagon tiles decorated by NiNi). 
Thus we suggest
   \begin{equation}
     \Htile = c_{72} N_{72} + \Esf N_{\rm sf}. 
     \label{eq-tileH}
   \end{equation}
where $N_{72}$ is the number of H(NiNi) tile pairs 
related by a $72^\circ$ rotation about their common tip, 
and $N_{\rm sf}$ is the total number of stacking flips.

We fitted $c_{72} =0.218$eV~\cite{alnico} from the simulation with
ideal sites; this is positive
since the $72^\circ$ relation of H(NiNi) tiles
creates intertile NiNi neighbors, and thus
reduces the number of (favorable) AlNi pairs. 
The structure minimizes $N_{72}$ by arranging that
in some places, an H(NiNi) tile has one tip
which contacts no other H(NiNi), and sees
CoAl (at a $144^\circ$ angle) in place of NiNi.  
That may be accomplished in two ways. 
If we re-allow lattice-gas hopping 
on the {\it interior} sites of the (small) HBS tiles, 
and properly adjust the stoichiometry~\cite{alnico}, 
we obtain the same old rigid decoration, 
except that a few H(CoAl) tiles appear, 
as illustrated in Fig.~\ref{fig-large}. 
On the other hand, if we retain the rigid NiNi decoration of small H tiles, 
the CoAl belongs to the interior of a Boat tile, and the large 
HBS tiling acquires a fourth kind of tile (``Bowtie''), 
as in Fig.~4(a) of Ref.~\onlinecite{alnico}.

Existing experiments can in principle measure $\Esf$. 
The time-dependent local ``phason'' flips~\cite{edagawa}, 
observed at $T\approx 1200$K by TEM through very thin samples of $d$(AlCuCo), 
can only occur by nucleating a ``stacking flip'' at one surface,
which subsequently random-walks to the other surface. 
Video images~\cite{edagawa} display an intermediate state for a
noticeable fraction -- say 10\% -- of the time, so that
with a sample thickness estimated at $100$~\AA~ (25 bilayers), 
one would very roughly estimate $\Esf\approx 0.3$ eV in $d$(AlCuCo). 
It would be exciting if such observations could be analyzed 
quantitatively.
In our simulation, with rigid  atoms decorating ideal sites, 
$ \Esf \approx 1.4 $ eV. However, relaxation 
(as in Fig.~\ref{fig-relaxed}) nearly cancels
this energy cost, indeed $\Esf \approx -0.2$ eV in very preliminary results, 
which should not be compared with the experiments on a different quasicrystal.
(Physically, $\Esf < 0$ probably implies a period 
$2c\approx 8$~\AA modulation.) 

\section{Discussion}

We find it remarkable that, with quite sketchy
experimental input, our simulations appear
competitive with single-crystal diffraction as a way of discovering
this rather complex structure.  This occurs despite the shortcomings
of the pair potentials -- our omission of 3-body terms, and the 
likelihood that the $V_{\rm AlTM}$ we use is stronger than the real one. 
Even if the connection to the microscopics is not quantitative, 
one may still obtain the large-scale order quite well:
that depends mostly on the tile Hamiltonian 
(\ref{eq-tileH}) having the proper form, which 
is more robust than the numerical values of the coefficients. 

It should also be pointed out that even imperfect potentials
could be quite useful for {\it augmenting} a fit of diffraction data. 
A combination of energy and diffraction data could overcome minor
spurious features in refinements~\cite{rtdiff}.

Our results strongly suggest that, in {\it this} alloy system, 
stability of the quasicrystal requires a ternary not only 
to tune the electron density, as in the Hume-Rothery picture, 
but also because each species fills a particular type of site. 
Thus it would be highly desirable to explore some other composition 
regimes in simulations. The Co-rich ``basic Co'' subphase of $d$(AlNiCo)
would be of particular interest. It exhibits much stronger 
diffuse scattering than Basic Ni, 
particularly halfway between the layers of Bragg peaks in reciprocal space, 
indicating a local tendency to doubling 
the $c$ periodicity ($8$~\AA~ structures). 

Our constraint that the atoms sit on ideal sites in the initial stage
of energy optimization has considerable potential to distort the conclusions.
Clearly, atom displacements have a very strong effect
so it is crucial to include these for a final answer. 
Small adjustments of position can gain as much energy as swaps of species, 
and in some cases the optimal sites may lie halfway between ideal sites, 
or between ideal layers -- the so-called ``puckered'' layers 
We do allow such relaxations in later stages, after determining 
a decoration, but it appears possible that a
different decoration would be obtained, if we allowed
relaxations in the early stages. 
This seems to be the most serious drawback of the
lattice-gas approach~\cite{cockayne} we used. 
Its advantages are that the systematic Monte Carlo
exploration is not biased by its users' prejudices, and
furthermore it could readily be adapted to $T>0$
(the temperatures at which the quasicrystal phase is stable). 
Nonzero temperature probably affects the result
just as strongly as relaxed positions do.

Within the framework of pair potentials and tile-decoration 
representation of the structure, a different  
approach is possible~\cite{decoII}, whereby relaxed sites are used in
the main discovery process. 
To implement this so that only valid tile decorations get explored, 
one performs relaxations with the decoration-equivalent atoms 
of a given class (``orbit'') being constrained to move together. 
However, that method is ad-hoc, depending on its 
users' educated guesses as to the tilings and decorations to be tried. 
Future work should strive to blend the best features of these two approaches 
to using decorations with potentials for the discovery of structures.

\thanks{This research was supported by the 
Department of Energy grant DE-FG02-89ER-45405
and by National Science Foundation grant DMR-0111198.
We thank E. Cockayne, F. G\"ahler,  E. Abe, and K. Edagawa
for discussions.}

\begin{thebibliography}{99}

\bibitem{alnico}
M.~Mihalkovi\v{c}, I.~Al-Lehyani, E.~Cockayne,
C~.L.~Henley, N.~Moghadam, J~.A.~Moriarty,
Y.~Wang, and M.~Widom, 
``Total-energy-based prediction of a quasicrystal structure'',
to appear, Phys. Rev. B; see www.arXiv.org, cond-mat/0102085. 

\bibitem{basicNi} 
S.~Ritsch,  C.~Beeli, H.~U.~Nissen, T. G\"odecke, M.~Scheffer, and
R.~L\"uck, 
{Phil. Mag. Lett.}, {74}, (1996) 99-106;
A.~P.~Tsai, A.~Fujiwara, A.~Inoue, and T.~Masumoto,  
{Phil. Mag. Lett.}, {74} (1996) 233-40.

\bibitem{PhaseDiagram} 
S.~Ritsch,  C.~Beeli, H.~U.~Nissen, T. G\"odecke, M.~Scheffer, and
R.~L\"uck, 
{Phil. Mag. Lett.} {78} (1998) 67-76.

\bibitem{takakura}
H. Takakura, A. Yamamoto, and A.-P. Tsai, 
Acta Cryst. A 57 (2001) 576-585. 

\bibitem{cervellino}
A. Cervellino, T.~Haibach, and W. Steurer, 
submitted to Acta Cryst. B, 2001. 

\bibitem{yan}
Y.~Yan,  S.~J.~Pennycook,  and A.~P.~Tsai,  
{Phys. Rev.  Lett.} {81} (1998) 5145-8.

\bibitem {abe}
E.~Abe, K.~Saitoh, H.~Takakura, A.~P.~Tsai, P.~J.~Steinhardt, and H.-C.~Jeong, 
Phys. Rev. Lett. 84 (2000) 4609-12. 

\bibitem{FN-NiCo}
Our Ni/Co assignments (based on potentials)
are completely different from those of Ref.~\onlinecite{abe}, 
which were guessed so as to match the site frequencies
in a simple, disorder-free decoration
with the experimental stoichiometry. 

\bibitem{Binary}
F. Lan\c con, L. Billard, and P.~Chaudhari, Europhys. Lett. {2}, (1986) 625;
M. Widom, K.~J.~Strandburg and R.~H.~Swendsen, 
Phys. Rev. Lett. {58} (1987) 706.  

\bibitem {dzugutov} 
    M. Dzugutov, Phys. Rev. Lett. {70} (1993) 2924.

\bibitem {rothDzu}
J. Roth and A.~R.~Denton, Phys. Rev. E 61 (2000) 6845-6857. 

\bibitem{krajci}
M.~Kraj\v c\'i,  J.~Hafner, and M.~Mihalkovi\v c,  
{Phys.  Rev. B} {62} (2000) 243-55.
The structure models used disagree with our 
present model in some important points. 

\bibitem{cockayne} 
E.~Cockayne, and M.~Widom, {\it Phys. Rev. Lett.} {81} (1998) 598-601.


\bibitem{GPT-I} 
J.~A.~Moriarty, and M.~Widom,  {Phys. Rev. B} {56} (1997) 7905-17.

\bibitem{GPT-II} 
M.~Widom and J.~A.~Moriarty, {Phys. Rev. B} {58} (1998) 8967-8979.

\bibitem{GPT-IV}
I.~Al-Lehyani, M.~Widom, Y.~Wang, N.~Moghadam, G.~M.~Stocks, and
J.~A.~Moriarty, 
{Phys.  Rev. B} 64 (2001) 075109.


\bibitem {decoII}
M.~Mihalkovi\v{c}, W.-J.~Zhu, C.~L.~Henley,
and R.~Phillips,
Phys. Rev. B 53 (1996) 9021-45.

\bibitem{FN-HumeR}  Stabilization due to Friedel oscillations is
{\it mathematically equivalent} to the Hume-Rothery
stabilization mechanism (if energies are computed in
second-order perturbation theory); the difference is that
one is expressed in real space and the other in reciprocal space. 


\bibitem{FN-whystable}
This rationale for structural tendencies emphasizes second-neighbor
TM-TM interactions, whereas
Ref.~\onlinecite{alnico}
emphasized first-neighbor Al-Al and Al-TM interactions
with Al atoms at ideal sites.
Perhaps the best ternary structures 
evade ``frustration'', by being independently
optimal for {\it both} TM-TM {\it and}
Al-related potentials. 


\bibitem{hen-decag}
C.~L.~Henley,
J. Non-Cryst. Solids 153\&154 (1993) 172-176.

\bibitem{FN-cockayne}
In particular, 
a similar lattice-gas annealing using potentials representing $d$(AlCuCo) 
obtained a somewhat different atomic structure.
(Ref.~\onlinecite{cockayne}.)
In contrast to the present formulation, 
the lattice-gas sites of Ref.~\onlinecite{cockayne} 
lay on a fixed quasilattice. 
The quasilattice sites must be more densely placed in order
to obtain comparable results without the freedom of tile flips. 


\bibitem{FN-fakering}
The 20~\AA~ rings are commoner in images than in our simulations
of single bilayers.  One possible explanation 
is ``stacking flips'' between the layers. 
This was illustrated in Fig.~8 of Ref.~\onlinecite{rtdiff},
using the same $d$(AlNiCo) structure reported in the present paper.
Just a couple stacking flips, seen in projection,
create an apparent decagon, and they also tend to symmetrize the 
appearance of the decagon interior. 

\bibitem{rtdiff}
C.~L.~Henley, V.~Elser, and M.~Mihalkovi\v{c},
Z. Kristallogr. 215 (2000) 553-568.

\bibitem{abe-phason}
Recent experiments which extract differences
in the TEM images of Basic Ni $d$(AlNiCo) samples upon rapid
temperature changes, indicate smearing of probability
density for the five Al atoms at the decagon  center, which is
reminiscent the distributions in our Fig.~\ref{fig-MDfuzz}. 
(E. Abe and A.~P.~Tsai, this conference.) 

\bibitem{gaehler-MD}
Independent MD simulations, using our structure  
and potentials (Refs.~\onlinecite{GPT-I,GPT-II,GPT-IV}), 
suggest that Al atoms can cyclically permute
in troughs around a Co atom or around the entire decagon.
(Franz G\"ahler, personal communication).

\bibitem{yan-calc}
Y.~Yan and S.~J.~Pennycook, {\it Phys. Rev. Lett.} {86} (2001) 1542-1545.


\bibitem{edagawa}
K. Edagawa, K.~Suzuki, and S.~Takeuchi, Phys. Rev. Lett. 
85 (2000) 1674, and this conference. 





\end{thebibliography}
\enddocument